\documentclass[prl,superscriptaddress,twocolumn]{revtex4}
\usepackage{graphicx}

\def\lno{La$_2$NiO$_4$}
\def\lsno{La$_{2-x}$Sr$_x$NiO$_4$}
\def\lsco{La$_{2-x}$Sr$_x$CuO$_4$}
\def\lnsco{La$_{1.6-x}$Nd$_{0.4}$Sr$_x$CuO$_4$}

\def\qco{${\bf q}_{\rm co}$}
\def\tco{$T_{\rm co}$}

\begin{document}
\title{Bond-Stretching-Phonon Anomalies in Stripe-Ordered
La$_{1.69}$Sr$_{0.31}$NiO$_4$} 
\author{J. M.~Tranquada}
\affiliation{Physics Department, Brookhaven National Laboratory, Upton,
NY 11973, USA}
\email{jtran@bnl.gov}
\author{K. Nakajima}
\affiliation{Neutron Scattering Laboratory, ISSP, University of
Tokyo, Tokai, Ibaraki, Japan}
\author{M. Braden}
\affiliation{Forschungszentrum Karlsruhe, IFP, Postfach 3640,
D-76021 Karlsruhe, Germany}
\affiliation{Laboratoire L\'eon Brillouin, CEA/CNRS, F-91191
Gif-sur-Yvette CEDEX, France}
\author{L. Pintschovius}
\affiliation{Forschungszentrum Karlsruhe, IFP, Postfach 3640,
D-76021 Karlsruhe, Germany}
\author{R. J. McQueeney}
\affiliation{Los Alamos National Laboratory, Los Alamos, NM 87545, USA}
\date{\today}
\begin{abstract}
We report a neutron scattering study of bond-stretching phonons in
La$_{1.69}$Sr$_{0.31}$NiO$_4$, a doped antiferromagnet in which the
added holes order in diagonal stripes at 45$^\circ$ to the Ni-O bonds. 
For the highest-energy longitudinal optical mode along the bonds,
a softening of 20\%\ is observed between the Brillouin zone center and
zone boundary.  At 45$^\circ$ to the bonds, a splitting of the same
magnitude is found across much of the zone.  Surprisingly, the
charge-ordering wave vector plays no apparent role in the anomalous
dispersions.  The implications for related anomalies in the cuprates are
discussed.
\end{abstract}
\maketitle

There is resurgent interest in the role of phonons
with respect to the high-temperature superconductivity found in layered
copper-oxides \cite{zhao97}.  Particularly
striking are the anomalies in high-energy optical modes observed by
neutron scattering in La$_{2-x}$Sr$_x$CuO$_4$
\cite{pint98,pint99,mcqu99} and YBa$_2$Cu$_3$O$_{6+x}$
\cite{pint98,reic96,petr00,mcqu01}.  There have been various speculations
as to whether the observed phonon anomalies might be related to
instantaneous charge inhomogeneities, particularly those in the form of
stripes \cite{emer99}.  Recently, detailed analyses of electron-phonon
interactions in a dimerized stripe phase have been reported
\cite{cast01,park01}.

One way to learn about the effect of charge stripes on lattice dynamics is
to study a model system with well-defined stripe order.  Here we present
the first single-crystal study, to our knowledge, of the bond-stretching
phonon modes in such a system, specifically La$_{2-x}$Sr$_x$NiO$_4$ with
$x=0.31$.  The stripe order in Sr-doped nickelates has been characterized
in detail by neutron diffraction, and the most recent summary of results
is given in
\cite{yosh00}.  For $x\gtrsim0.22$, the crystal structure is tetragonal,
consisting of a body-centered stacking of NiO$_2$ planes.  Within the
NiO$_2$ planes, the charge stripes run diagonally along either [110] or
$[1\bar{1}0]$ directions, at 45$^\circ$ to the Ni-O nearest-neighbor
bonds, which extend along [100] and [010] directions.  (In contrast, the
charge stripes observed in superconducting \lnsco\ run parallel to the
Cu-O bonds \cite{ichi00}; however, for $x\lesssim0.06$, the stripes
inferred to exist in \lsco\ have the diagonal orientation of the
nickelates \cite{waki99,fuji01}.)  The maximum transition temperatures
for charge-stripe and magnetic order occur at $x=\frac13$
\cite{yosh00,cheo94}.  Well below the charge-ordering temperature, \tco,
the nickelates have very large resistivities \cite{kats96,kats99},
consistent with all of the added holes being localized in charge stripes.

We focus on the highest-energy longitudinal optical modes propagating
along the [100] and [110] directions.  These modes show only weak
dispersion in stoichiometric \lno\ \cite{pint01}.  In contrast, we
observe a softening of 20\%\ along [100] on moving from the Brillouin zone
center to the zone boundary, quite similar to that found in \lsco\
\cite{pint99,mcqu99}; along [110], a splitting of the same magnitude is
observed over much of the zone and, in particular, at the zone boundary. 
These results are important for two reasons.  1) The observed anomalies
must be associated with the local charge inhomogeneity.  They are induced
by the hole doping, and the holes are localized in the stripes.  2) 
There is no evidence that the charge-ordering wave vector, \qco, plays a
special role.  If the phonon anomalies were related to collective phase
fluctuations of the charge stripes, as in a conventional
charge-density-wave system \cite{schu78}, then one might expect them to
appear at \qco.  With the absence of a collective signature, it seems
likely that the dominant effects involve local interactions between
charge and lattice fluctuations.

Our \lsno\ crystal, grown by the floating-zone method, is cylindrical,
with a diameter of 6 mm and length of 30 mm; the Sr concentration of
$x=0.31$ was confirmed by inductively-coupled plasma analysis (with an
uncertainty of $\pm0.01$). At room temperature, the lattice parameters of
the tetragonal unit cell are $a=3.84$~\AA\ and
$c=12.70$~\AA.   The charge and spin ordering transitions were confirmed
to be consistent with previous work\cite{yosh00} (magnetic and
charge-order transitions of approximately 160~K and 235~K, respectively)
by neutron diffraction measurements performed at the JRR-3M reactor in
Tokai, Japan.  The inelastic-neutron-scattering measurements of phonons
were performed on the 1T triple-axis spectrometer at the Orphe\'e reactor
of the Laboratoire L\'eon Brillouin in Saclay, France.  The spectrometer
is equipped with a Cu (111) monochromator and a pyrolytic graphite (PG)
(002) analyzer, each of which is both vertically and horizontally
focused.  The analyzer was set to detect neutrons with a final frequency
of 3.55 THz (14.7 meV).  A PG filter was placed after the sample to
minimize unwanted neutrons at harmonic wavelengths.

Figure 1(a) shows scans of scattered intensity versus
excitation frequency for phonons propagating in the [100] direction,
parallel to the in-plane Ni-O bonds.  The scans are taken at wave vectors
along the line
${\bf q}=(h,0,0)$ in the first Brillouin zone, moving from the zone
center ($h=0$, bottom) to the zone boundary ($h=1$, top), where the
components of the wave vector are measured in reciprocal lattice units,
$(2\pi/a,2\pi/a,2\pi/c)$.   (The effective zone boundary for a single
NiO$_2$ plane is at $h=0.5$; the fact that the actual boundary is
at $h=1$ results from the body-centered stacking of the layers.)   Each
scan is dominated by a single, well-defined peak, that disperses from
approximately 21 THz at zone center to less than 18 THz half-way across
the zone, and then back up again.  The lines through the data points are
fitted gaussian peaks on top of a background that is taken to be
independent of wave vector and frequency.

\begin{figure}[t]
\centerline{\includegraphics[width=3.4in]{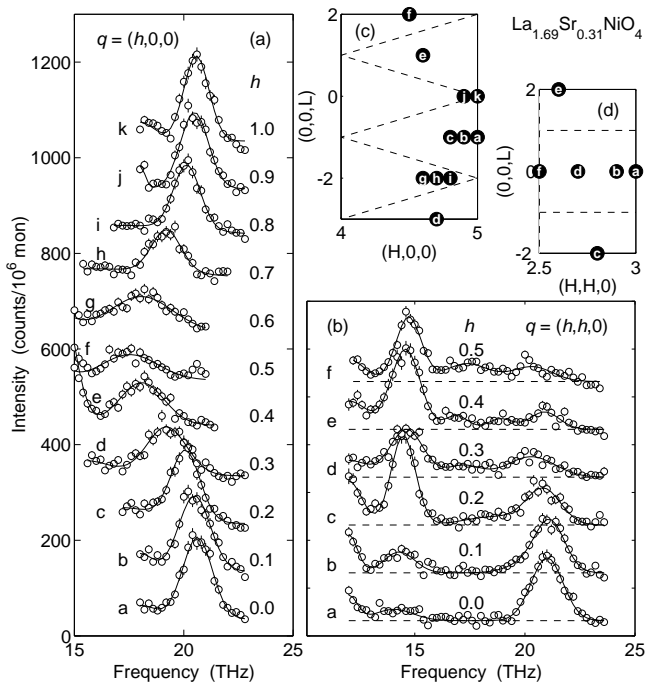}}
\medskip
\caption{Neutron scattering measurements of phonons in
La$_{1.69}$Sr$_{0.31}$NiO$_4$ at $T=10$~K.  (a) Scattered intensity vs.\
frequency for reduced wave vectors, ${\bf q} = (h,0,0)$, varying from
Brillouin zone center ($h=0$) to the zone boundary $(h=1)$.  Curves
through the data are fits to gaussian peak shapes on top of a constant
background.  Scans have been offset vertically for clarity.  (b) Similar
results for
${\bf q}=(h,h,0)$; dashed lines indicate the constant background.  (c) and
(d) indicate the positions in reciprocal space where the scans
(identified by letters) of (a) and (b), respectively, were measured; 
dashed lines indicate Brillouin zone boundaries.  Measurements were
performed in various equivalent zones, with varying $L$-component, in
order to avoid spurious peaks due to accidental Bragg scattering by the
sample.}
\label{fg:scans}
\end{figure}

Figure 1(b) shows scans measured along the [110] direction.  Because of
twinning of the stripe domains, we simultaneously probe phonons
propagating parallel and perpendicular to the stripes.  Again, we focus
on the mode that starts at about 21 THz at zone center; the mode at 14
THz involves bond-bending motion, and does not exhibit any notable
doping-dependent behavior.  [The intensity variation seen in the figure
for this mode is associated with its sensitivity to the $L$-component of
the wave vector {\bf Q}, which varies from point to point as noted in
Fig.~1(d).] On moving from zone center to zone boundary, we note that the
highest-frequency mode varies little in frequency, but it loses
intensity.  Concommitant with this, signal appears in the 17--18 THz
regime, growing in strength as the zone boundary is approached.   

Using the gaussian fits as a smoothed version of the data, we present our
results as intensity plots in Fig.~2(a) and (b).  Here, the intensity has
been multiplied by $\nu/Q_\|^2$ ($Q_\|$ is the component of {\bf Q} along
the phonon propagation direction) to correct for sensitivities of the
neutron-scattering cross section.  Dramatic differences are observed in
the dispersions of the stripe-ordered system compared to those in
stoichiometric \lno, which are indicated by the gray circles
\cite{pint01}.  Contrary to naive expectations based on the
Peierls-distortion model \cite{schu78}, there is no particular anomaly at
the charge-ordering wave vector
${\bf q}_{\rm CO}=(0.31,0.31,0)$ [see Fig.~2(b)]. Instead, we find a
splitting of the mode along [110] over a substantial part of the zone.   
The size of the dispersion in the [100] direction is about the same as
the [110] splitting, and is essentially identical to the magnitude of the
dopant-induced softening reported for superconducting
La$_{1.85}$Sr$_{0.15}$CuO$_4$ \cite{pint99,mcqu99}.  The softened modes
are consistent with a previous study of the phonon density-of-states in
\lsno\ \cite{mcqu99b}, where a dopant-induced peak was observed at
$\sim75$ meV ($\approx18$~THz).

\begin{figure}[t]
\centerline{\includegraphics[width=3.4in]{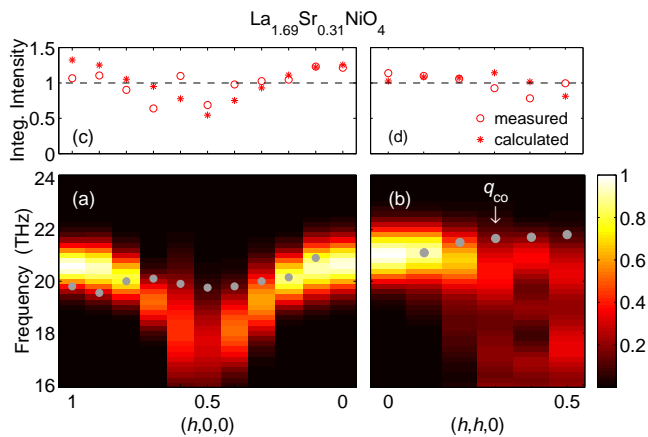}}
\medskip
\caption{(color) False-color images of intensity as a function of
frequency and
${\bf q}$ for (a) ${\bf q}=(h,0,0)$ and (b) ${\bf q}=(h,h,0)$.  The
plotted intensity corresponds to the fitted curves (minus the background)
from Fig.~2. $q_{\rm CO}$ indicates the approximate charge-ordering wave
vector. The gray circles indicate the phonon dispersions measured in
stoichiometric \lno \protect\cite{pint01}.  (c) Intensity vs.
${\bf q}$ integrated over frequency along the [100] direction.  Open
circles are data, stars are intensities calculated from the
interatomic-potential model for La$_{1.9}$NiO$_{3.93}$
\protect\cite{pint01}.  (d) Similar plot for the [110] direction. } 
\label{fg:images}
\end{figure}

Integrating the corrected intensity data of Fig.~2(a) and (b) over
frequency yields the open circles shown in Fig.~2(c) and (d),
respectively.  The experimental results are compared with intensities
calculated from the interatomic-potential model used to describe
the phonons in La$_{1.9}$NiO$_{3.93}$ \cite{pint01}.  (An overall
normalization of the calculated to the measured intensities has been
applied in each panel.) The general consistency between the measured and
calculated integrated intensities suggests that the distribution of
weight for the anomalously-softened bond-stretching modes does not
extend significantly below 16~THz.

To interpret our results, let us consider some simple phenomenological
models.  The high-energy bond-stretching phonons involve the motion of
the in-plane oxygens between the much heavier nickel atoms.  Suppose we
assume the nickel atoms to be infinitely heavy and consider only
nearest-neighbor forces; then each oxygen acts like an Einstein
oscillator, with no dispersion of its vibrational frequency.  Correcting
for the finite mass of the nickel would result in a slight decrease in
frequency towards the zone boundary; however, accounting for Coulomb
repulsion between neighboring oxygen ions would counter that with an
increase in frequency at the zone boundary, where neighboring oxygens
move opposite to one another.  Ignoring these corrections, a simple
Einstein model gives a rough approximation of the measured dispersions in
stoichiometric \lno\ \cite{pint01} [see Fig.~2(a) and (b)].

\begin{figure}[t]
\centerline{\includegraphics[width=3.4in]{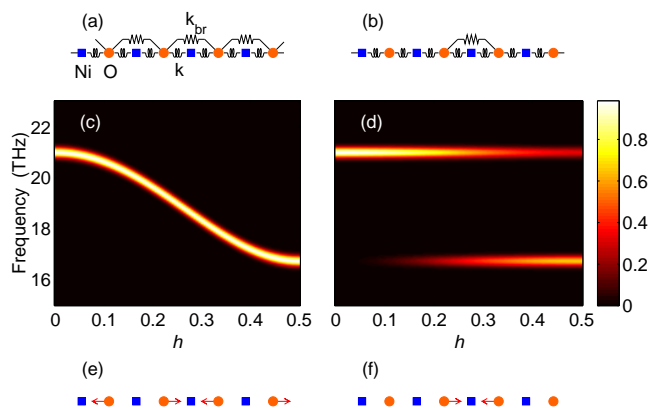}}
\medskip
\caption{(color) (a) Sketch of a force-constant model for a linear Ni-O
array with a nearest-neigbor force constant of strength $k$, and an
effective force constant $k_{\rm br}$ between next-nearest-neighbor
oxygens.  (b) Model with $k_{\rm br}$ acting only between one out of
every three oxygen pairs.  (c) Calculated intensity vs.\
$q=(2\pi/a)h$ and frequency for the model in (a) with $(2k/m_{\rm
O})^{1/2}/2\pi=21$~THz and 
$(-2k_{\rm br}/m_{\rm O})^{1/2}/2\pi=9$~THz.  (d) Calculated
intensity for model (b) with $(2k/m_{\rm O})^{1/2}/2\pi=21$~THz and 
$(-k_{\rm br}/m_{\rm O})^{1/2}/2\pi=9$~THz.  (e) Sketch of the eigenmode
at $h=0.5$ for model (a), corresponding to a linear breathing mode.  (f) 
Softened eigenmode at $h=0.5$ for model (b), corresponding to a local
breathing mode.} 
\label{fg:models}
\end{figure}

To go further, we will restrict ourselves to a linear Ni-O chain model. 
The softening of phonons at the zone boundary can be described by
introducing a force between nearest-neighbor O ions with a negative
force constant, $k_{\rm br}$ \cite{pint89,pint01} [see
Fig.~\ref{fg:models}(a)].  This force phenomenologically incorporates
a particular type of electron-phonon coupling.   An example of the
dispersion from such a model is shown in Fig.~\ref{fg:models}(c).  The
eigenvector for the softened zone-boundary mode, sketched in
Fig.~\ref{fg:models}(e), involves linear breathing motion of the O about
the Ni.  

Next, suppose that $k_{\rm br}$ acts only between one pair of oxygens out
of three [see Fig.~\ref{fg:models}(b)].  The result is a splitting of the
modes, as illustrated in Fig.~\ref{fg:models}(d).  Note that the
intensity of the softened mode is strong at the zone boundary, but goes
to zero at zone center.  This effect is achieved without a dimerization
of the lattice.  As shown in Fig.~\ref{fg:models}(f), the eigenvector of
the softened mode at zone boundary involves an isolated, local breathing
motion of a pair of O about a Ni site.  The behavior of this mode does
not require coherence between motions in different unit cells.

It seems likely that both of these models have some degree of relevance
to our observations; however, it is difficult to associate them in a
consistent or unique way with specific details of the measured
dispersions.  Of greater interest is the nature of the electron-phonon
coupling modelled by $k_{\rm br}$.  In model calculations for \lsco,
Falter and Hoffmann \cite{falt01} have emphasized the importance of ionic
charge fluctuations in response to atomic displacements and the
irrelevance of Fermi-surface nesting effects of the type once proposed by
Weber \cite{webe87}.  Given the experimental evidence for poor electronic
screening of phonons in optimally-doped cuprates \cite{home00}, extended
Coulomb interactions should also be important.

In the present case of the nickelate, little metallic screening is
expected due to the gap of $>0.1$~eV in the optical conductivity at low
temperature \cite{kats96,pash00}.  The bond-stretching modes involve polar
fluctuations of the negative oxygen ions against the positive nickel
ions, and so should couple to charge fluctuations.  The lowest-energy
channel for charge fluctuations must involve the dopant-induced holes,
and since the holes are segregated into stripes, charge fluctuations must
be associated with some form of stripe fluctuations.

In a conventional charge-density-wave (CDW) system, one might expect the
mode along [110] to couple to fluctuations of the phase of the CDW with
respect to the lattice \cite{schu78}.  In the present case, the absence of
a characteristic wave vector suggests that local charge fluctuations may
be more relevant for the bond-stretching modes.  Local fluctuations were
predicted by Yi {\it et al.} \cite{yi98} in a random-phase-approximation
treatment of Hartree-Fock stripes.  Nevertheless, our results do not rule
out the possibility of weak, low-energy, collective phase fluctuations of
the stripes.

\begin{figure}[t]
\centerline{\includegraphics[width=2.8in]{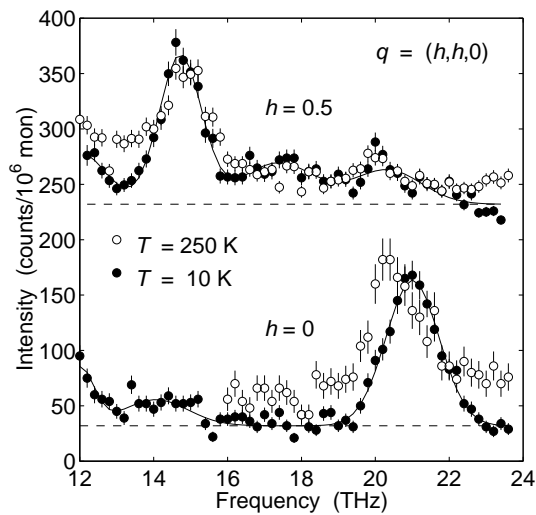}}
\medskip
\caption{Comparison of scans for ${\bf q}=(h,h,0)$ at temperatures of
10~K (filled circles) and 250~K (open circles).  The lines are the same as
in Fig.~2.  The upper scans have been shifted vertically by 200 counts.} 
\label{fg:temp}
\end{figure}

Do the effects we observe depend on static ordering of the stripes? 
Figure~\ref{fg:temp} shows a comparison of phonons at zone-center and
zone boundary along [110] measured at 10~K and 250~K, which is above the
charge-ordering temperature.  One can see that at 250~K, the zone-center
mode has softened slightly but remains strong and well-defined; however,
at the zone boundary, the intensity remains spread out in frequency. 
Thus, the anomalies do not disappear with the static order.

Rather similar phonon softenings were observed in the sample identified as
La$_{1.9}$NiO$_{3.93}$ \cite{pint01,pint89}.  Despite the specified
stoichiometry, that sample exhibited superlattice peaks of the type
$(H/2,K/2,L/2)$ with $H$, $K$, and $L$ odd, consistent with the stage-2
ordering of interstitials found in La$_2$NiO$_{4.105}$ \cite{tran94b}. 
The latter compound exhibits evidence of incipient stripe correlations,
but no stripe order \cite{tran97c}.  This is further evidence that local
interactions are sufficient to yield the phonon anomalies.

To summarize, we have observed doping-induced anomalies in
bond-stretching modes of La$_{1.69}$Sr$_{0.31}$NiO$_4$, a
compound in which the doped holes order in stripes at low temperature. 
The anomalies persist over a substantial portion of the Brillouin zone,
with no obvious signature of collective stripe fluctuations.   The
splitting of the mode along [110] is similar to recent observations in
superconducting Ba$_{0.6}$K$_{0.4}$BiO$_3$
\cite{brad01}, where charge inhomogeneity is also expected to be
relevant.   These
results provide support for a connection between the phonon anomalies
observed in the cuprates and instantaneous charge inhomogeneity.

We gratefully acknowledge stimulating discussions with R. Werner and
helpful comments from P. D. Johnson and S. A. Kivelson. This work was
supported by the Materials Sciences Division, Office of Science, U.S.
Department of Energy under Contract No.\ DE-AC02-98CH10886, and by the
U.S.-Japan Cooperative Research Program on Neutron Scattering.  JMT and
KN wish to thank the staff of the Laboratoire L\'eon Brillouin for their
hospitality during the experiments.


\end{document}